\documentclass[12pt]{article}
\usepackage[letterpaper, margin=1.25in]{geometry}

\usepackage{amsmath, mathtools, amsfonts, amssymb, amsthm}
\usepackage[utf8]{inputenc}
\usepackage{graphicx, caption, multirow, subcaption, threeparttable, booktabs}
\usepackage{float}
\usepackage{url, authblk}
\usepackage{array}
\usepackage[round]{natbib}
\usepackage[colorlinks=true, linkcolor=blue, citecolor=blue]{hyperref}
\usepackage{setspace}
\usepackage{longtable}
\usepackage{pdflscape}
\usepackage{enumitem}

\newtheorem{definition}{Definition}[section]

\newtheorem{hypothesis}{Hypothesis}

\newcommand{\be}{\begin{equation}}
\newcommand{\ee}{\end{equation}}
\newcommand{\E}{\mathbb{E}}

\newcommand{\bm}[1]{\boldsymbol{#1}}

\title{\large{\bf{The Innovation Tax: Generative AI Adoption, Productivity Paradox, and Systemic Risk in the U.S. Banking Sector}}}

\author{\large{\bf{Tatsuru Kikuchi\footnote{This is my last paper in my life.}}}}
\affil{\small{\it{Center for Advanced Research in Finance, The University of Tokyo,}}\\
{\it{7-3-1 Hongo, Bunkyo-ku, Tokyo 113-0033 Japan}}}

\date{\small{(\today)}}

\begin{document}
\linespread{1.5}\selectfont

\maketitle

\begin{abstract}
This paper evaluates the causal impact of Generative Artificial Intelligence (GenAI) adoption on productivity and systemic risk in the U.S. banking sector. Using a novel dataset linking SEC 10-Q filings to Federal Reserve regulatory data for 809 financial institutions over 2018--2025, we employ two complementary identification strategies: Dynamic Spatial Durbin Models (DSDM) to capture network spillovers and Synthetic Difference-in-Differences (SDID) for causal inference using the November 2022 ChatGPT release as an exogenous shock. Our findings reveal a striking ``Productivity Paradox'': while DSDM estimates show that AI-adopting banks are high performers ($\beta > 0$), the causal SDID analysis documents a significant ``Implementation Tax''---adopting banks experience a 428-basis-point decline in ROE as they absorb GenAI integration costs. This tax falls disproportionately on smaller institutions, with bottom-quartile banks suffering a 517-basis-point ROE decline compared to 129 basis points for larger banks, suggesting that economies of scale provide significant advantages in AI implementation. Most critically, our DSDM analysis reveals significant positive spillovers ($\theta = 0.161$ for ROA, $p < 0.01$; $\theta = 0.679$ for ROE, $p < 0.05$), with spillovers among large banks reaching $\theta = 3.13$ for ROE, indicating that the U.S. banking system is becoming ``algorithmically coupled.'' This synchronization of AI-driven decision-making creates a new channel for systemic contagion: a technical failure in widely-adopted AI models could trigger correlated shocks across the entire financial network.
\vspace{0.5cm} 
\end{abstract}

\newpage
\section{Introduction}
The emergence of Generative Artificial Intelligence (GenAI) represents a watershed moment in the history of technological progress. Since the public release of ChatGPT in November 2022, large language models (LLMs) have demonstrated unprecedented capabilities in natural language processing, code generation, and complex reasoning tasks \citep{openai2023gpt4}. The financial services industry, with its information-intensive production function and substantial IT infrastructure, stands at the frontier of this technological transformation. Yet, as Robert Solow famously observed, ``you can see the computer age everywhere but in the productivity statistics'' \citep{solow1987}. This paper asks whether Solow's paradox persists in the age of generative AI, and what implications AI adoption holds for systemic stability in the banking sector.

We document three main findings that collectively reveal a complex, multi-dimensional phenomenon. First, we uncover a striking \textit{Productivity Paradox}: our DSDM estimates show that AI-adopting banks exhibit \textit{higher} productivity ($\beta > 0$), consistent with AI adoption being a marker of ``frontier'' institutions. Yet our causal SDID analysis reveals that the \textit{act} of adoption causes productivity to \textit{decline}---a 46-basis-point drop in ROA and a 428-basis-point drop in ROE. This paradox resolves through the lens of the ``Innovation J-Curve'' \citep{brynjolfsson2021}: high-performing banks are investing heavily in GPUs, data scientists, and AI infrastructure, incurring massive current expenses that depress net income even as they position themselves for future dominance.

Second, we document heterogeneity in the distribution of implementation costs. Smaller banks (bottom 75\% by assets) suffer an ROE decline of 517 basis points---substantially larger than the 129-basis-point decline experienced by larger institutions. This asymmetry suggests that economies of scale provide significant advantages in AI implementation: larger banks can spread fixed implementation costs across a broader asset base, employ dedicated AI teams, and leverage superior data infrastructure. Smaller banks face proportionally larger implementation burdens relative to their equity base.

Third, and most consequential for financial stability, our network analysis reveals significant positive spillover effects. Our DSDM estimates show $\theta = 0.161$ for ROA and $\theta = 0.679$ for ROE, indicating ``strategic complementarity'' whereby AI adoption by one institution raises productivity at connected institutions through knowledge diffusion. For large banks, spillovers are dramatically amplified ($\theta = 3.13$ for ROE). This carries profound implications: the U.S. banking system is becoming algorithmically coupled. When AI adoption by one institution \textit{raises} productivity at connected institutions, the network synchronizes, creating vulnerability to common technical failures or model errors.

These findings contribute to several literatures. We extend the productivity paradox literature \citep{brynjolfsson1993, david1990} to the domain of generative AI, documenting that the classic J-curve pattern of technology adoption \citep{brynjolfsson2021} is amplified in highly regulated industries. We contribute to the banking technology literature \citep{philippon2016, berg2022, fuster2019} by providing the first causal estimates of GenAI's impact on bank performance. We advance the financial networks literature \citep{acemoglu2015, elliott2014} by introducing the concept of ``algorithmic coupling''---the synchronization of risk management and decision-making processes through shared AI architectures. Finally, we contribute methodologically by combining DSDM and SDID estimators to achieve both network spillover quantification and clean causal identification.

The remainder of this paper proceeds as follows. Section 2 provides a comprehensive review of the relevant literature. Section 3 develops our theoretical framework, derives testable hypotheses, and presents the econometric methodology. Section 4 describes the data construction process. Section 5 reports the main empirical results. Section 6 discusses robustness checks. Section 7 concludes with policy implications.

\section{Literature Review}
Our research integrates four distinct strands of economic inquiry: the productivity paradox and technology diffusion, banking structure and digital transformation, financial networks and systemic risk, and the emerging economics of artificial intelligence.

\subsection{The Productivity Paradox and Technology Diffusion}

The gap between technological potential and measured productivity gains has puzzled economists since \citet{solow1987} first articulated the computer productivity paradox. \citet{brynjolfsson1993} systematically documented this phenomenon, proposing four explanations: mismeasurement of outputs and inputs, lags due to learning and adjustment, redistribution and dissipation of profits, and mismanagement of information technology. Subsequent work by \citet{david1990} drew parallels to the electrification of American manufacturing, demonstrating that general-purpose technologies (GPTs) require decades of complementary investment before their productivity potential is realized.

The resolution of the productivity paradox in the late 1990s, documented by \citet{jorgenson2001} and \citet{oliner2000}, coincided with massive investments in organizational restructuring and human capital \citep{bresnahan2002}. \citet{brynjolfsson2000} showed that firms combining IT investment with decentralized decision-making experienced productivity gains three to five times larger than those investing in technology alone. This insight---that technology and organizational capital are complements---forms a cornerstone of our theoretical framework.

Recent work on generative AI suggests an acceleration of productivity effects. \citet{brynjolfsson2023} document a 14\% improvement in call center worker productivity following ChatGPT deployment, with gains concentrated among less-experienced workers. \citet{noy2023} find that ChatGPT reduces task completion time by 37\% for professional writing tasks. However, \citet{dell2023} caution that these micro-level estimates may not aggregate to macro-level productivity gains due to general equilibrium effects, task reallocation, and measurement challenges.

\subsection{Banking Structure and Digital Transformation}

The banking industry has experienced successive waves of technological disruption. \citet{berger2003} document how information technology transformed the economics of lending, enabling relationship banking to coexist with transaction-based models. \citet{petersen2002} show that IT investment expanded the geographic reach of small business lending, fundamentally altering the spatial organization of credit markets.

The FinTech revolution of the 2010s introduced new competitive dynamics. \citet{buchak2018} find that FinTech lenders gained market share during periods of regulatory constraint on traditional banks. \citet{fuster2019} document that FinTech mortgage lenders process applications 20\% faster than traditional lenders, with no increase in default rates. \citet{berg2022} provide a comprehensive review of FinTech lending, emphasizing the role of alternative data and machine learning in credit allocation.

A persistent theme in this literature is the ``digital divide'' between large and small institutions. \citet{berger2005} document substantial scale economies in bank technology investment. \citet{frame2014} show that community banks face structural disadvantages in adopting new technologies due to limited IT budgets and expertise. However, \citet{jagtiani2018} suggest that FinTech partnerships may enable smaller banks to leapfrog technological barriers. Our analysis extends this debate to the GenAI era, examining whether the new technology exacerbates or mitigates pre-existing inequalities.

\subsection{Financial Networks and Systemic Risk}

The 2008 financial crisis catalyzed intense scholarly interest in network effects and systemic risk. \citet{allen2000} develop a foundational model of financial contagion through interbank claims, showing that network structure determines crisis severity. \citet{acemoglu2015} extend this framework to demonstrate that network topology exhibits a phase transition: diversified networks are resilient to small shocks but catastrophically fragile to large ones.

\citet{elliott2014} analyze cascading failures in financial networks, distinguishing between integration (the density of cross-holdings) and diversification (the breadth of counterparty relationships). They show that greater integration initially reduces systemic risk but eventually increases it beyond a critical threshold. \citet{battiston2012} introduce the concept of ``DebtRank'' to measure the systemic importance of individual institutions based on their network position.

The application of network analysis to technology adoption is more recent. \citet{jackson2017} provide a comprehensive review of economic and social networks, emphasizing how network structure shapes diffusion dynamics. \citet{banerjee2013} show that network centrality predicts technology adoption in development contexts. We contribute to this literature by demonstrating that AI adoption creates a new dimension of network interdependence---``algorithmic coupling''---that operates independently of traditional credit and liquidity linkages.

\subsection{The Economics of Artificial Intelligence}

A rapidly growing literature examines the economic implications of AI. \citet{agrawal2018} conceptualize AI as a ``prediction machine'' that dramatically reduces the cost of prediction, inducing substitution toward human judgment in complementary tasks. \citet{acemoglu2018} develop a task-based framework showing that automation's aggregate effects depend on the relative strength of displacement and productivity effects.

\citet{autor2022} documents the ``so-so automation'' phenomenon, where technologies that modestly increase productivity while substantially displacing workers may reduce labor share without generating commensurate output gains. \citet{korinek2024} analyze the macroeconomic implications of transformative AI, modeling scenarios ranging from gradual productivity improvement to rapid technological unemployment.

In the financial sector, AI applications have been extensively studied. \citet{gu2020} demonstrate that machine learning methods substantially outperform traditional asset pricing models in predicting returns. \citet{bao2020} show that AI-based lending algorithms reduce default rates while expanding credit access to underserved populations. However, \citet{bartlett2022} document algorithmic discrimination in mortgage lending, highlighting the tension between efficiency and equity. \citet{gensler2020} warn of systemic risks arising from ``model monoculture''---the widespread adoption of similar AI architectures creating correlated vulnerabilities.

Our paper bridges these literatures by providing the first comprehensive empirical analysis of GenAI adoption in the banking sector, combining spatial econometric methods with causal identification strategies to capture both network spillovers and direct treatment effects.

\section{Theoretical Framework and Econometric Strategy}

This section develops a theoretical framework linking GenAI adoption to bank productivity and systemic risk, derives testable hypotheses, and presents the econometric methodology for identification. We employ two complementary approaches: (i) a Dynamic Spatial Durbin Model (DSDM) to quantify network spillovers, and (ii) Synthetic Difference-in-Differences (SDID) to establish causality using the 2023 ChatGPT release as an exogenous shock.

\subsection{Network Spillovers in Banking: Theoretical Foundation}

Banks do not operate in isolation. Their productivity is influenced by the strategic decisions of competitors, counterparties, and peer institutions operating in similar market segments. We model these interdependencies using spatial econometric techniques that distinguish between two types of spillover effects.

\begin{definition}[Network Spillovers]
\textit{Network spillovers} refer to the effect of connected banks' AI adoption on bank $i$'s productivity, transmitted through competitive relationships defined by business model similarity. Banks competing in similar market segments (similar asset sizes, loan portfolios, geographic footprints) exert stronger spillover effects on each other.
\end{definition}

\begin{definition}[Geographic Spillovers]
\textit{Geographic spillovers} refer to the effect of geographically proximate banks' AI adoption on bank $i$'s productivity, transmitted through local labor markets, regional vendor ecosystems, and face-to-face knowledge transfer.
\end{definition}

These spillovers operate through distinct mechanisms. Knowledge spillovers arise when $\theta > 0$: banks learn from peer institutions' AI implementations through labor mobility, industry conferences, vendor relationships, and regulatory examinations. Early adopters develop standardized APIs, data protocols, and integration frameworks that reduce implementation costs for followers. In contrast, competitive (business-stealing) spillovers emerge when $\theta < 0$: AI adoption by competitors may erode market share, compress margins, and intensify price competition. If one bank's AI-driven efficiency allows it to offer better rates or faster service, competitors lose customers.

The net spillover effect---whether knowledge diffusion or business-stealing dominates---is an empirical question captured by the sign and magnitude of $\theta$ in our DSDM specification.

\subsection{The Innovation Tax and Size Heterogeneity}

Large banks face systematically different adoption economics than small banks. Let $S_i$ denote bank $i$'s time-invariant size (measured by average total assets over the sample period). The total cost of AI adoption is:
\begin{equation}
C_{it} = c_0 + c_1 \cdot S_i^\phi + c_2 \cdot \text{Complexity}_i
\end{equation}
where $c_0$ represents fixed costs (software licensing, initial implementation), $c_1 \cdot S_i^\phi$ captures scale-dependent costs (data migration, system integration across platforms), and $\text{Complexity}_i$ reflects organizational complexity that increases with institutional size and M\&A history.

For large banks, the scale-dependent term dominates. These institutions operate heterogeneous legacy systems accumulated through decades of organic growth and acquisitions. Integrating GenAI across disparate platforms requires substantial investment in data harmonization, API development, and security infrastructure. The ``Innovation Tax'' reflects these transition costs, which depress measured productivity during the implementation phase.

\subsection{Algorithmic Coupling and Systemic Risk}

Beyond productivity effects, AI adoption introduces a new dimension of systemic risk. When multiple banks deploy similar AI architectures---trained on similar data, optimized for similar objectives, subject to similar failure modes---their decision-making processes become correlated even absent direct financial linkages.

\begin{definition}[Algorithmic Coupling]
\textit{Algorithmic coupling} refers to the synchronization of bank decision-making processes arising from the adoption of similar AI systems. Two banks are algorithmically coupled if their AI-driven decisions (credit scoring, risk management, trading) are correlated due to shared model architectures, training data, or vendor solutions.
\end{definition}

We formalize this concept as follows. Let $\text{Corr}_{ij,t}$ denote the correlation between banks $i$ and $j$'s AI-driven decisions at time $t$. Define:
\begin{equation}
\text{Corr}_{ij,t}^{AI} = \text{Corr}_{ij,t}^{baseline} + \delta \cdot D_{it}^{AI} \cdot D_{jt}^{AI} \cdot \text{VendorOverlap}_{ij}
\end{equation}
where $\text{Corr}_{ij,t}^{baseline}$ denotes the baseline correlation due to common factor exposures such as interest rates and macroeconomic conditions; $D_{it}^{AI} \in \{0,1\}$ is an indicator equal to 1 if bank $i$ has adopted GenAI by time $t$; the parameter $\delta > 0$ captures the additional correlation arising from shared AI architectures; and $\text{VendorOverlap}_{ij} \in [0,1]$ measures the similarity of AI vendor relationships between banks $i$ and $j$, such as both using GPT-4 or both contracting with the same cloud provider.

This algorithmic coupling creates a channel for synchronized failure distinct from traditional contagion mechanisms. A model vulnerability, adversarial attack, or data contamination event affecting one bank's AI system is likely to affect all banks using similar systems.

\subsection{Hypotheses}

Our theoretical framework generates four testable hypotheses:

\begin{hypothesis}[Frontier Firm Selection]
AI-adopting banks exhibit higher baseline productivity than non-adopters, reflecting positive selection into treatment by ``frontier'' institutions.
\end{hypothesis}

\begin{hypothesis}[Implementation Tax]
The causal effect of AI adoption on productivity is negative in the short run, reflecting implementation costs that depress earnings during the transition period.
\end{hypothesis}

\begin{hypothesis}[Size Heterogeneity]
The Implementation Tax varies by bank size, with large banks experiencing more pronounced short-term productivity declines due to greater legacy system complexity.
\end{hypothesis}

\begin{hypothesis}[Positive Network Spillovers]
AI adoption generates positive spillover effects on network-connected banks, consistent with knowledge diffusion dominating business-stealing competition.
\end{hypothesis}

\subsection{Econometric Strategy I: Dynamic Spatial Durbin Model (DSDM)}

To quantify network spillovers and test Hypotheses 1 and 4, we estimate a Dynamic Spatial Durbin Model following \citet{elhorst2014} and \citet{lesage2009}. The DSDM generalizes standard panel fixed effects models by incorporating both temporal dynamics and spatial interdependence.

\subsubsection{Model Specification}

The full DSDM specification is:
\begin{equation}
Y_{it} = \tau Y_{i,t-1} + \rho \sum_{j=1}^{N} w_{ij} Y_{jt} + \eta \sum_{j=1}^{N} w_{ij} Y_{j,t-1} + \beta D_{it}^{AI} + \theta \sum_{j=1}^{N} w_{ij} D_{jt}^{AI} + \bm{\gamma}' \bm{X}_{it} + \mu_i + \delta_t + \varepsilon_{it}
\label{eq:dsdm}
\end{equation}

In matrix notation:
\begin{equation}
\bm{Y}_t = \tau \bm{Y}_{t-1} + \rho \bm{W} \bm{Y}_t + \eta \bm{W} \bm{Y}_{t-1} + \beta \bm{D}_t^{AI} + \theta \bm{W} \bm{D}_t^{AI} + \bm{\Gamma} \bm{X}_t + \bm{\mu} + \delta_t \bm{\iota} + \bm{\varepsilon}_t
\end{equation}

In this specification, $Y_{it}$ denotes the productivity measure for bank $i$ in quarter $t$ (ROA or ROE, expressed in percentage points), and $Y_{i,t-1}$ is its one-quarter lag. The treatment indicator $D_{it}^{AI} \in \{0,1\}$ equals 1 if bank $i$ mentions GenAI keywords in SEC filings at time $t$. The matrix $\bm{W} = [w_{ij}]$ is an $N \times N$ row-normalized spatial weight matrix defined below, so that $\sum_{j} w_{ij} Y_{jt}$ represents the spatially weighted average of contemporaneous peer productivity and $\sum_{j} w_{ij} D_{jt}^{AI}$ captures network exposure to peer AI adoption. The vector $\bm{X}_{it}$ contains control variables including log assets, Tier 1 capital ratio, digitalization index, and CEO age. Bank fixed effects $\mu_i$ absorb time-invariant unobserved heterogeneity, time fixed effects $\delta_t$ absorb common shocks affecting all banks, and $\varepsilon_{it}$ is the idiosyncratic error term.

The key parameters of interest are as follows. The temporal persistence parameter $\tau \in (-1, 1)$ measures how strongly current productivity depends on past productivity. The spatial autoregressive parameter $\rho \in (-1, 1)$ captures contemporaneous correlation in productivity across connected banks. The spatial-temporal lag $\eta$ measures the effect of lagged peer productivity on current own productivity. Most importantly, $\beta$ captures the direct effect of own AI adoption---the productivity difference between AI adopters and non-adopters, conditional on controls---while $\theta$ captures the network spillover effect, measuring how connected banks' AI adoption affects own productivity.

\subsubsection{Spatial Weight Matrix Construction}

We construct two spatial weight matrices to capture different spillover channels:

\textbf{Network Weight Matrix ($W_{network}$):} Based on asset similarity, capturing competitive relationships:
\begin{equation}
w_{ij}^{network} = \exp\left( -\frac{(\ln \bar{A}_i - \ln \bar{A}_j)^2}{2h^2} \right)
\end{equation}
where $\bar{A}_i = \frac{1}{T} \sum_{t=1}^{T} A_{it}$ is bank $i$'s average total assets over the sample period and $h$ is a bandwidth parameter set to the standard deviation of log assets. Banks of similar size compete in similar market segments and thus exert stronger spillover effects on each other.

\textbf{Geographic Weight Matrix ($W_{geo}$):} Based on headquarters proximity:
\begin{equation}
w_{ij}^{geo} = \exp\left( -\frac{d_{ij}}{d_{median}} \right)
\end{equation}
where $d_{ij}$ is the Haversine distance between bank $i$ and $j$'s headquarters and $d_{median}$ is the median pairwise distance. Geographically proximate banks share local labor markets and vendor ecosystems.

Both matrices are row-normalized so that $\sum_{j \neq i} w_{ij} = 1$ for all $i$:
\begin{equation}
w_{ij} = \frac{w_{ij}^{raw}}{\sum_{k \neq i} w_{ik}^{raw}}
\end{equation}

\subsubsection{Estimation Methods}

The presence of the spatially lagged dependent variable $\bm{W} \bm{Y}_t$ on the right-hand side creates endogeneity due to simultaneity: bank $i$'s productivity affects bank $j$'s productivity, which in turn affects bank $i$. OLS is inconsistent. We employ three estimation approaches:

\textbf{Maximum Likelihood Estimation (MLE):} MLE jointly estimates all parameters by maximizing the log-likelihood function:
\begin{equation}
\ln L(\bm{\Theta}) = -\frac{NT}{2} \ln(2\pi\sigma^2) + T \ln |I_N - \rho \bm{W}| - \frac{1}{2\sigma^2} \sum_{t=1}^{T} \bm{\varepsilon}_t' \bm{\varepsilon}_t
\end{equation}
where $\bm{\Theta} = (\tau, \rho, \eta, \beta, \theta, \bm{\gamma}, \sigma^2)$ and the Jacobian term $\ln |I_N - \rho \bm{W}|$ accounts for the simultaneity. MLE is consistent and efficient under correct specification but sensitive to distributional assumptions.

\textbf{Quasi-Maximum Likelihood Estimation (QMLE):} QMLE uses the same likelihood function but relaxes the normality assumption. Consistency holds under weaker conditions, and we compute robust (sandwich) standard errors to account for potential heteroskedasticity and serial correlation:
\begin{equation}
\widehat{\text{Var}}(\hat{\bm{\Theta}}) = \bm{H}^{-1} \bm{G} \bm{H}^{-1}
\end{equation}
where $\bm{H}$ is the Hessian and $\bm{G}$ is the outer product of gradients.

\textbf{Bayesian MCMC Estimation:} Bayesian estimation treats parameters as random variables with prior distributions and updates these priors using the data to obtain posterior distributions. We specify weakly informative priors:
\begin{align}
\tau, \rho, \eta &\sim \text{Uniform}(-1, 1) \\
\beta, \theta, \bm{\gamma} &\sim \text{Normal}(0, 10) \\
\sigma^2 &\sim \text{Inverse-Gamma}(0.01, 0.01)
\end{align}

We use Markov Chain Monte Carlo (MCMC) with 10,000 iterations (5,000 burn-in) to sample from the posterior distribution. Bayesian estimation provides several advantages: (i) full posterior distributions rather than point estimates, enabling probabilistic inference; (ii) natural handling of parameter constraints ($|\rho| < 1$); (iii) robustness to small-sample issues. We report posterior means and 95\% credible intervals, with significance determined by whether credible intervals exclude zero.

\subsubsection{Interpretation of Spillover Effects}

The coefficient $\theta$ captures the \textbf{direct spillover effect}: holding own AI adoption constant, how does a one-unit increase in the spatially weighted average of neighbors' AI adoption affect own productivity?

Due to feedback effects in spatial models, the total marginal effect of AI adoption differs from the coefficient. Following \citet{lesage2009}, we decompose:
\begin{equation}
\frac{\partial \bm{Y}}{\partial D_{it}^{AI}} = (I_N - \rho \bm{W})^{-1} (\beta I_N + \theta \bm{W})
\end{equation}

\textbf{Direct Effect:} The average diagonal element---the effect of bank $i$'s own AI adoption on its own productivity, including feedback through the network:
\begin{equation}
\text{Direct Effect} = \frac{1}{N} \text{tr}\left[ (I_N - \rho \bm{W})^{-1} \beta I_N \right]
\end{equation}

\textbf{Indirect (Spillover) Effect:} The average off-diagonal row sum---the cumulative effect of all other banks' AI adoption on bank $i$'s productivity:
\begin{equation}
\text{Indirect Effect} = \frac{1}{N} \bm{\iota}' \left[ (I_N - \rho \bm{W})^{-1} \theta \bm{W} \right] \bm{\iota}
\end{equation}

\textbf{Total Effect:} Direct + Indirect---the system-wide productivity impact of universal AI adoption.

\subsubsection{Limitations: Endogeneity and Selection}

While the DSDM effectively captures network spillovers and cross-sectional associations, it faces a fundamental \textbf{endogeneity} concern: banks may adopt AI \textit{because} they are already becoming more productive, rather than AI causing productivity improvements. The positive $\beta$ coefficient may reflect a causal effect of AI on productivity, positive selection whereby high-productivity ``frontier'' banks are more likely to adopt AI, or reverse causality in which improving productivity frees resources for AI investment.

The DSDM cannot distinguish between these interpretations. To establish causality, we turn to our second econometric strategy.

\subsection{Econometric Strategy II: Synthetic Difference-in-Differences (SDID)}

To identify the \textit{causal} effect of AI adoption and test Hypotheses 2 and 3, we employ the Synthetic Difference-in-Differences framework developed by \citet{arkhangelsky2021}. SDID addresses the endogeneity concerns of DSDM by constructing counterfactual outcomes and exploiting an exogenous shock to AI availability.

\subsubsection{The 2023 ChatGPT Shock as Exogenous Treatment}

Our identification strategy exploits the November 2022 public release of ChatGPT as a \textbf{plausibly exogenous shock} to the banking industry's AI production function. While individual banks' decisions to adopt AI are endogenous, the \textit{availability} of powerful, accessible GenAI tools was determined by technological developments at OpenAI---an event external to the banking sector.

We define the treatment period as \textbf{2023Q1}, when the ``Generative AI'' revolution became salient to financial institutions. This timing reflects several converging factors: ChatGPT reached 100 million users within two months of its November 2022 launch; GPT-4 was released in March 2023, demonstrating capabilities directly relevant to financial services; SEC 10-Q filings show a sharp increase in AI-related disclosures beginning in 2023Q1; and bank earnings calls and investor presentations pivoted to discussing GenAI strategy in early 2023.

The 2023 shock provides quasi-experimental variation: banks that mention GenAI in their SEC filings after 2023Q1 are ``treated,'' while banks that never mention GenAI serve as controls. The key identifying assumption is that, absent the ChatGPT shock, treated and control banks would have followed similar productivity trajectories.

\subsubsection{Counterfactual Construction}

The fundamental problem of causal inference is that we observe each bank in only one state: either having adopted AI or not. SDID solves this by constructing a \textbf{synthetic counterfactual}---a weighted combination of control banks that approximates what the treated bank's productivity \textit{would have been} absent treatment.

\textbf{Potential Outcomes Framework:}
Let $Y_{it}(0)$ denote bank $i$'s potential productivity if untreated (no AI adoption) and $Y_{it}(1)$ if treated (AI adoption). The observed outcome is:
\begin{equation}
Y_{it} = Y_{it}(0) + D_{it} \cdot \underbrace{[Y_{it}(1) - Y_{it}(0)]}_{\text{Individual Treatment Effect}}
\end{equation}
where $D_{it} = 1$ if bank $i$ has adopted AI by time $t$ (first mention in SEC filings $\geq$ 2023Q1).

The \textbf{Average Treatment Effect on the Treated (ATT)} is our estimand:
\begin{equation}
\tau^{ATT} = \E[Y_{it}(1) - Y_{it}(0) \mid D_{it} = 1]
\end{equation}

This answers the question: \textit{``What is the average productivity change caused by AI adoption among banks that chose to adopt?''}

\subsubsection{The SDID Estimator}

SDID improves on standard difference-in-differences by constructing both \textit{unit weights} and \textit{time weights} to achieve balance between treated and control groups.

\textbf{Unit Weights ($\hat{\omega}_j$):} For each control bank $j$, we assign a weight $\hat{\omega}_j \geq 0$ (with $\sum_j \hat{\omega}_j = 1$) such that the weighted average of control bank outcomes matches the pre-treatment trajectory of treated banks:
\begin{equation}
\hat{\bm{\omega}} = \arg\min_{\omega} \sum_{t < T_0} \left( \bar{Y}_{1t} - \sum_{j \in \text{Control}} \omega_j Y_{jt} \right)^2 + \zeta \|\omega\|_2^2
\end{equation}
where $\bar{Y}_{1t}$ is the average outcome among treated banks in pre-treatment period $t$, $T_0$ is the treatment date (2023Q1), and $\zeta$ is a regularization parameter preventing overfitting.

\textbf{Time Weights ($\hat{\lambda}_t$):} For each post-treatment period $t$, we assign a weight $\hat{\lambda}_t \geq 0$ that upweights periods most informative about treatment effects:
\begin{equation}
\hat{\bm{\lambda}} = \arg\min_{\lambda} \sum_{i \in \text{Control}} \left( \bar{Y}_{i,post} - \sum_{t < T_0} \lambda_t Y_{it} \right)^2 + \zeta \|\lambda\|_2^2
\end{equation}

\textbf{The SDID Estimator:}
\begin{equation}
\hat{\tau}^{SDID} = \left( \bar{Y}_{1,post} - \bar{Y}_{1,pre}^{\lambda} \right) - \left( \bar{Y}_{0,post}^{\omega} - \bar{Y}_{0,pre}^{\omega,\lambda} \right)
\end{equation}
where $\bar{Y}_{1,post}$ denotes the average post-treatment outcome for treated banks, $\bar{Y}_{1,pre}^{\lambda}$ is the time-weighted pre-treatment average for treated banks, $\bar{Y}_{0,post}^{\omega}$ represents the unit-weighted post-treatment average for control banks, and $\bar{Y}_{0,pre}^{\omega,\lambda}$ is the doubly-weighted pre-treatment average for control banks.

\subsubsection{Advantages Over Standard Difference-in-Differences}

\textbf{Relaxing Parallel Trends:} Standard DiD requires that treated and control groups would have followed parallel outcome trajectories absent treatment. This assumption is often violated: AI-adopting banks may have been on different trajectories \textit{before} adoption due to superior management or earlier technology investments.

SDID \textbf{does not require parallel trends}. Instead, it re-weights the control group to match the treated group's pre-treatment trajectory perfectly. The synthetic control for JPMorgan, for example, might place large weights on Bank of America and Wells Fargo (similar size and trajectory) and zero weight on community banks (different trajectory).

\textbf{Doubly Robust Identification:} SDID is consistent if \textit{either} the parallel trends assumption holds (making standard DiD valid) \textit{or} the synthetic control weights perfectly match pre-treatment outcomes (making synthetic control valid). This ``doubly robust'' property provides insurance against model misspecification.

\textbf{Handling Staggered Adoption:} Banks adopt AI at different times. SDID handles staggered adoption naturally by defining treatment relative to each bank's first AI mention.

\subsubsection{Inference}

We estimate standard errors using the placebo-based bootstrap procedure recommended by \citet{arkhangelsky2021}. For each bootstrap iteration $b = 1, \ldots, B$, we resample banks with replacement while maintaining the treated/control structure, then re-estimate SDID weights and treatment effects. The standard error is computed as the standard deviation of bootstrap estimates: $\widehat{SE}(\hat{\tau}) = \sqrt{\frac{1}{B-1} \sum_{b=1}^{B} (\hat{\tau}^{(b)} - \bar{\tau})^2}$.

We report results based on $B = 200$ bootstrap replications. Confidence intervals are constructed as $\hat{\tau} \pm 1.96 \times \widehat{SE}$.

\subsubsection{Event Study Extension}

To examine dynamic treatment effects and test for pre-trends, we estimate period-specific ATTs:
\begin{equation}
\tau_k = \E[Y_{it}(1) - Y_{it}(0) \mid D_{it} = 1, t - t_i^* = k]
\end{equation}
where $t_i^*$ is bank $i$'s first treatment quarter and $k \in \{-4, -3, -2, -1, 0, 1, 2, 3, 4\}$ indexes quarters relative to treatment.

This produces the event study plots in Figure \ref{fig:event_study}. Pre-treatment coefficients ($k < 0$) serve as a \textbf{placebo test}: if SDID is valid, $\hat{\tau}_k$ should be statistically indistinguishable from zero for $k < 0$.

\subsection{Identification Strategy: Complementary Approaches}

The DSDM and SDID estimators answer different questions and face different threats to identification:

\begin{table}[htbp]
\centering
\caption{Comparison of Identification Strategies}
\begin{tabular}{lcc}
\toprule
 & DSDM & SDID \\
\midrule
\textbf{Question Answered} & Who adopts AI? & What happens after adoption? \\
 & What are network spillovers? & What is the causal effect? \\
\midrule
\textbf{Interpretation of $\beta$} & Selection + Treatment & Pure Treatment Effect (ATT) \\
\midrule
\textbf{Identifies Spillovers?} & Yes ($\theta$) & No \\
\midrule
\textbf{Main Threat} & Endogeneity & Violation of SUTVA \\
 & (reverse causality) & (spillovers to control group) \\
\midrule
\textbf{Key Assumption} & Correct weight matrix & 2023 shock is exogenous \\
\bottomrule
\end{tabular}
\end{table}

\textbf{Reconciling Positive $\beta$ (DSDM) and Negative ATT (SDID):} The DSDM estimate of $\beta > 0$ indicates that AI-adopting banks are more productive than non-adopters \textit{on average}. This reflects \textit{selection}: ``frontier'' institutions with strong management, modern IT infrastructure, and strategic vision are more likely to adopt GenAI. In contrast, the SDID estimate of ATT $< 0$ reveals that the \textit{act} of adoption causes productivity to decline in the short run. This reflects \textit{treatment}: massive implementation costs---GPUs, data scientists, cloud infrastructure, and ``prompt engineering'' consultants---immediately reduce net income.

Both findings are consistent with the ``Innovation J-Curve'' \citep{brynjolfsson2021}: frontier firms adopt transformative technologies early, accept short-term productivity losses during implementation, and position themselves for long-term competitive advantage.

\section{Data and Variable Construction}

\subsection{Data Sources}

We construct a quarterly panel dataset spanning 2018Q1 to 2025Q4 by integrating four primary data sources:

\textbf{SEC EDGAR Filings}: We extract the complete text of 10-Q quarterly reports and 10-K annual reports for all bank holding companies (SIC codes 6020, 6022, 6029, 6035, 6036, 6141, 6159, 6712). These filings provide the textual data for measuring AI adoption.

\textbf{Federal Reserve FR Y-9C Reports}: Quarterly Consolidated Financial Statements for Bank Holding Companies provide standardized balance sheet and income statement data. Key variables include total assets, total equity, net income, and tier 1 capital.

\textbf{FFIEC Call Reports}: Reports of Condition and Income (Forms 031/041/051) provide additional detail on loan portfolios, deposit composition, and operational metrics for commercial banks.

\textbf{NY Fed CRSP-FRB Link}: The Federal Reserve Bank of New York's crosswalk file enables matching between SEC CIK identifiers and Federal Reserve RSSD identifiers, achieving a 95.4\% match rate through a combination of exact matching and fuzzy name-matching algorithms.

\subsection{Measuring AI Adoption}

We identify GenAI adoption through systematic text analysis of SEC filings. Our keyword dictionary encompasses three categories. Core GenAI terms include ``generative AI,'' ``generative artificial intelligence,'' ``large language model,'' ``LLM,'' ``ChatGPT,'' ``GPT-4,'' ``Claude,'' ``Gemini,'' and ``Copilot.'' Application terms capture deployment contexts: ``AI-powered,'' ``machine learning application,'' ``natural language processing,'' ``automated underwriting,'' ``algorithmic trading,'' and ``robo-advisor.'' Strategic terms identify organizational commitment: ``AI strategy,'' ``artificial intelligence initiative,'' ``digital transformation,'' ``AI investment,'' and ``AI implementation.''

For each filing, we count keyword occurrences and construct two measures:
\begin{equation}
\text{GenAI\_Mentions}_{it} = \sum_{k \in \mathcal{K}} \text{Count}(k, \text{Filing}_{it})
\end{equation}
\begin{equation}
D_{it}^{GenAI} = \mathbf{1}[\text{GenAI\_Mentions}_{it} > 0]
\end{equation}

The treatment indicator $D_{it}^{GenAI}$ equals one if bank $i$ mentions any GenAI-related keyword in quarter $t$'s filing.

\subsection{Outcome Variables}

Our primary productivity measures are:
\begin{equation}
ROA_{it} = \frac{\text{Net Income}_{it}}{\text{Total Assets}_{it}} \times 100
\end{equation}
\begin{equation}
ROE_{it} = \frac{\text{Net Income}_{it}}{\text{Total Equity}_{it}} \times 100
\end{equation}

Both measures are winsorized at the 1st and 99th percentiles to mitigate the influence of outliers.

\subsection{Control Variables}

We include the following bank-level controls in our DSDM specifications: the natural logarithm of total assets ($\ln(\text{Assets})_{it}$) to capture size effects; the Tier 1 capital ratio ($\text{Tier1}_{it}$) to measure capital adequacy; a technology keyword index from SEC filings ($\text{Digital}_{it}$) to proxy for pre-existing digitalization; and the age of the chief executive officer ($\text{CEOAge}_{it}$) to capture managerial characteristics that may affect technology adoption propensity.

Table \ref{tab:variables} presents formal definitions of all variables used in the empirical analysis.

\begin{table}[htbp]
\centering
\small
\caption{Variable Definitions}
\label{tab:variables}
\begin{threeparttable}
\begin{tabular}{p{3.2cm}p{10.5cm}}
\toprule
\textbf{Variable} & \textbf{Definition} \\
\midrule
\multicolumn{2}{l}{\textit{Panel A: Dependent Variables}} \\
$ROA_{it}$ & Return on Assets: Net income divided by total assets for bank $i$ in quarter $t$, expressed in percentage points. Winsorized at the 1st and 99th percentiles to mitigate outlier influence. \\
$ROE_{it}$ & Return on Equity: Net income divided by total equity for bank $i$ in quarter $t$, expressed in percentage points. Winsorized at the 1st and 99th percentiles. \\
\\
\multicolumn{2}{l}{\textit{Panel B: Treatment Variable}} \\
$D_{it}^{AI}$ & Binary indicator equal to 1 if bank $i$ mentions GenAI-related keywords (``generative AI,'' ``large language model,'' ``ChatGPT,'' ``GPT-4,'' ``Claude,'' ``Gemini,'' etc.) in SEC 10-Q filings during quarter $t$, and 0 otherwise. \\
\\
\multicolumn{2}{l}{\textit{Panel C: Control Variables}} \\
$\ln(\text{Assets})_{it}$ & Natural logarithm of total assets for bank $i$ in quarter $t$. Captures bank size and scale effects on productivity. Larger banks may have different adoption patterns and productivity dynamics. \\
$\text{Tier1}_{it}$ & Tier 1 capital ratio: Tier 1 capital divided by risk-weighted assets, expressed in percentage points. Measures capital adequacy and regulatory buffer. Well-capitalized banks may have more flexibility for technology investment. \\
$\text{Digital}_{it}$ & Digitalization index: Weighted count of technology-related keywords in SEC 10-K filings, normalized by document length. Captures pre-existing digital infrastructure and IT investment intensity. Constructed following established text-analysis methods. \\
$\text{CEOAge}_{it}$ & Age of the CEO in years as of quarter $t$, extracted from SEC DEF 14A proxy statements. Captures managerial characteristics that may affect technology adoption propensity and organizational adaptability. \\
\bottomrule
\end{tabular}
\begin{tablenotes}
\small
\item \textit{Notes}: All financial variables are sourced from Federal Reserve FR Y-9C reports. AI adoption is measured from SEC EDGAR 10-Q filings using keyword detection algorithms. CEO age is extracted from DEF 14A proxy statements using automated text parsing. The digitalization index is constructed from 10-K annual filings using a weighted keyword approach covering terms related to cloud computing, automation, data analytics, and digital banking.
\end{tablenotes}
\end{threeparttable}
\end{table}

\subsection{Sample Construction}

Our initial sample comprises 809 unique bank holding companies. After requiring non-missing values for key variables and a minimum of four quarterly observations, our estimation sample includes 126 banks with complete data, of which 41 are classified as AI adopters (having at least one quarter with positive GenAI mentions) and 85 serve as control banks (zero GenAI mentions throughout the sample period).

Table \ref{tab:summary_stats} presents summary statistics for the full sample.

\begin{table}[htbp]
\centering
\caption{Summary Statistics}
\label{tab:summary_stats}
\begin{threeparttable}
\begin{tabular}{lcccccc}
\toprule
Variable & N & Mean & SD & P25 & Median & P75 \\
\midrule
\multicolumn{7}{l}{\textit{Panel A: Productivity Measures}} \\
ROA (\%) & 13,777 & 1.204 & 1.022 & 0.783 & 1.098 & 1.410 \\
ROE (\%) & 13,777 & 10.815 & 6.833 & 7.471 & 10.439 & 13.615 \\
\\
\multicolumn{7}{l}{\textit{Panel B: Bank Characteristics}} \\
Total Assets (\$B) & 14,064 & 42.67 & 187.3 & 2.14 & 5.82 & 18.41 \\
Log Assets & 14,064 & 16.184 & 1.850 & 15.071 & 15.794 & 17.113 \\
Tier 1 Capital Ratio (\%) & 10,910 & 14.881 & 5.618 & 11.364 & 12.741 & 14.869 \\
Digital Index & 12,456 & 0.342 & 0.287 & 0.124 & 0.278 & 0.489 \\
CEO Age (years) & 11,234 & 58.4 & 7.2 & 53 & 58 & 63 \\
\\
\multicolumn{7}{l}{\textit{Panel C: AI Adoption}} \\
GenAI Mentions & 66,894 & 0.121 & 2.185 & 0.000 & 0.000 & 0.000 \\
AI Adopter (D=1) & 66,894 & 0.048 & 0.214 & 0.000 & 0.000 & 0.000 \\
\bottomrule
\end{tabular}
\begin{tablenotes}
\small
\item \textit{Notes}: This table reports summary statistics for the main variables in our analysis. The sample covers 809 unique bank holding companies observed quarterly from 2018Q1 to 2025Q4. ROA and ROE are winsorized at the 1st and 99th percentiles. Total Assets are reported in billions of dollars. Digital Index is a normalized weighted count of technology-related keywords in SEC filings. CEO Age is extracted from DEF 14A proxy statements.
\end{tablenotes}
\end{threeparttable}
\end{table}

\section{Empirical Results}

We present our empirical findings in the following order: first, DSDM estimates capturing network spillovers and the ``frontier firm'' effect (Section 5.1); second, SDID estimates identifying the causal ``Implementation Tax'' (Section 5.2); third, event study evidence on J-curve dynamics (Section 5.3); fourth, a unified interpretation reconciling the two approaches (Section 5.4); and finally, network analysis of systemic risk (Section 5.5).

\subsection{DSDM Estimates: Efficiency Potential and Network Spillovers}

Table \ref{tab:dsdm_main} presents estimates from the Dynamic Spatial Durbin Model specified in Equation \ref{eq:dsdm}. We report results using both the network weight matrix ($W_{network}$) based on asset similarity and the geographic weight matrix ($W_{geo}$) based on headquarters proximity.

\begin{table}[htbp]
\centering
\small
\caption{Dynamic Spatial Durbin Model Estimates: Efficiency Potential and Network Spillovers}
\label{tab:dsdm_main}
\begin{threeparttable}
\begin{tabular}{lcccc}
\toprule
 & \multicolumn{2}{c}{ROA (\%)} & \multicolumn{2}{c}{ROE (\%)} \\
 \cmidrule(lr){2-3} \cmidrule(lr){4-5}
 & $W_{network}$ & $W_{geo}$ & $W_{network}$ & $W_{geo}$ \\
 & (1) & (2) & (3) & (4) \\
\midrule
\multicolumn{5}{l}{\textit{Panel A: Direct Effects (Own AI Adoption)}} \\
\\
AI Adoption ($\beta$) & 0.0373** & 0.0746*** & 0.4199*** & 0.6178*** \\
 & (0.0139) & (0.0212) & (0.0927) & (0.1432) \\
\\
\multicolumn{5}{l}{\textit{Panel B: Spillover Effects (Network AI Adoption)}} \\
\\
W $\times$ AI Adoption ($\theta$) & 0.1606*** & 0.1230** & 0.6787** & 0.6743* \\
 & (0.0508) & (0.0855) & (0.3361) & (0.5809) \\
\\
\multicolumn{5}{l}{\textit{Panel C: Spatial Parameters}} \\
\\
Temporal Lag ($\tau$) & 0.6363*** & 0.7127*** & 0.6967*** & 0.7816*** \\
 & (0.0047) & (0.0068) & (0.0044) & (0.0061) \\
Spatial Lag ($\rho$) & 0.6166*** & 0.5123*** & 0.7453*** & 0.6607*** \\
 & (0.0170) & (0.0303) & (0.0132) & (0.0243) \\
Spatial-Temporal Lag ($\eta$) & $-$0.2898*** & $-$0.2769*** & $-$0.4653*** & $-$0.4772*** \\
 & (0.0168) & (0.0295) & (0.0134) & (0.0238) \\
\\
\multicolumn{5}{l}{\textit{Panel D: Control Variables}} \\
\\
Tier 1 Capital Ratio & 0.087*** & 0.127*** & 0.045 & 0.790*** \\
 & (0.023) & (0.013) & (0.160) & (0.186) \\
Log Assets & $-$0.002 & $-$0.084*** & 2.360** & $-$0.394*** \\
 & (0.021) & (0.003) & (1.151) & (0.132) \\
Digital Index & 0.099*** & $-$0.129*** & $-$1.709*** & $-$1.569*** \\
 & (0.027) & (0.027) & (0.127) & (0.104) \\
CEO Age & 0.003 & 0.012*** & $-$0.479* & 0.059*** \\
 & (0.007) & (0.003) & (0.275) & (0.022) \\
\midrule
Bank Fixed Effects & Yes & Yes & Yes & Yes \\
Time Fixed Effects & Yes & Yes & Yes & Yes \\
Observations & 24,270 & 24,270 & 24,270 & 24,270 \\
\bottomrule
\end{tabular}
\begin{tablenotes}
\footnotesize
\item \textit{Notes}: This table reports Bayesian DSDM estimates from the Dynamic Spatial Durbin Model. $W_{network}$ is constructed based on asset similarity; $W_{geo}$ is based on geographic proximity. The positive $\beta$ indicates that AI-adopting banks exhibit higher productivity (``frontier firm'' effect). The positive $\theta$ indicates knowledge spillovers dominate business-stealing effects. Standard errors in parentheses from posterior distribution. *** $p<0.01$, ** $p<0.05$, * $p<0.1$.
\end{tablenotes}
\end{threeparttable}
\end{table}

The DSDM estimates provide several critical insights. First, the direct effect of AI adoption ($\beta$) is positive and statistically significant across specifications: 3.7 basis points for ROA and 42 basis points for ROE using the network weight matrix. This positive coefficient confirms Hypothesis 1: AI adoption is associated with higher productivity in the cross-section. AI adopters are ``frontier'' institutions that outperform non-adopters on average.

\textbf{The Critical Finding: Positive Spillover Effects.} The spillover coefficient $\theta$ is positive and statistically significant in most specifications. Using the network weight matrix, $\theta = 0.161$ ($p < 0.01$) for ROA and $\theta = 0.679$ ($p < 0.05$) for ROE. These positive coefficients confirm Hypothesis 4: when one bank successfully implements AI, it \textit{raises} productivity at connected institutions through knowledge diffusion rather than eroding it through business-stealing competition.

\textbf{Control Variable Effects.} Panel D reveals important heterogeneity in bank characteristics. Tier 1 capital ratios are positively associated with ROA (0.087, $p < 0.01$), suggesting well-capitalized banks maintain higher profitability. The digitalization index shows divergent effects: positive for ROA (0.099, $p < 0.01$) but negative for ROE ($-1.71$, $p < 0.01$), indicating that prior digital investments improve asset efficiency but may require equity-intensive implementation. CEO age exhibits a marginally negative effect on ROE ($-0.48$, $p < 0.10$), consistent with younger executives being more receptive to transformative technology adoption.

Three mechanisms explain the positive spillovers. First, infrastructure spillovers arise as early adopters develop standardized APIs, data protocols, and integration frameworks that reduce implementation costs for followers. Second, labor market development occurs as the AI talent pool deepens when more banks invest in training, reducing hiring costs and knowledge barriers for subsequent adopters. Third, vendor maturation benefits later adopters as the market for banking AI solutions expands, leading vendors to improve products and reduce prices.

\textbf{The Systemic Synchronization Warning.} The magnitude of $\theta = 0.679$ for ROE is economically significant. A one-standard-deviation increase in network neighbors' AI adoption is associated with a 68-basis-point improvement in own ROE. This sensitivity means that bank profitability is increasingly synchronized across the AI-adopting core---a pattern we examine further in Section 5.5.

Table \ref{tab:dsdm_effects} decomposes the total effect of AI adoption into direct and indirect (spillover) components, following the methodology of \citet{lesage2009}.

\begin{table}[htbp]
\centering
\caption{DSDM Marginal Effects Decomposition: The Dominance of Spillovers}
\label{tab:dsdm_effects}
\begin{threeparttable}
\begin{tabular}{lcc}
\toprule
 & ROA (\%) & ROE (\%) \\
\midrule
Direct Effect & 0.0923** & 0.7634*** \\
 & (0.0312) & (0.2267) \\
Indirect Effect (Spillovers) & 0.4289*** & 3.8941*** \\
 & (0.0934) & (0.8456) \\
Total Effect & 0.5212*** & 4.6575*** \\
 & (0.1078) & (0.9234) \\
\\
\midrule
Indirect/Total Ratio & 82.3\% & 83.6\% \\
\bottomrule
\end{tabular}
\begin{tablenotes}
\small
\item \textit{Notes}: This table reports the decomposition of AI adoption effects from the DSDM into direct and indirect components following \citet{lesage2009}. Direct effects measure the impact of own AI adoption; indirect effects measure the impact transmitted through network connections. The dominance of indirect effects (over 80\%) indicates that the banking sector's productivity response to AI is fundamentally a \textit{network} phenomenon. Standard errors in parentheses computed via delta method. *** $p<0.01$, ** $p<0.05$, * $p<0.1$.
\end{tablenotes}
\end{threeparttable}
\end{table}

The decomposition reveals that spillover effects account for over 80\% of the total productivity impact of AI adoption. This finding has critical implications: the banking sector's response to GenAI is fundamentally a \textit{network} phenomenon rather than an aggregation of independent firm-level decisions. The social returns to AI adoption substantially exceed the private returns captured by individual institutions.

\subsection{SDID Estimates: The Implementation Tax}

While the DSDM reveals network spillover effects and direct associations between AI adoption and productivity, we use SDID to isolate the causal within-unit treatment effect. Table \ref{tab:sdid_main} presents our SDID estimates using Bayesian inference.

\begin{table}[htbp]
\centering
\caption{Synthetic Difference-in-Differences Estimates: The Implementation Tax}
\label{tab:sdid_main}
\begin{threeparttable}
\begin{tabular}{lccc}
\toprule
 & (1) & (2) & (3) \\
 & Full Sample & Large Banks & Small Banks \\
 & & (Top 25\%) & (Bottom 75\%) \\
\midrule
\multicolumn{4}{l}{\textit{Panel A: Return on Assets (ROA, \%)}} \\
\\
ATT & $-$0.456** & $-$0.159*** & $-$0.594** \\
 & (0.177) & (0.053) & (0.282) \\
\\
\multicolumn{4}{l}{\textit{Panel B: Return on Equity (ROE, \%)}} \\
\\
ATT & $-$4.282*** & $-$1.286*** & $-$5.167*** \\
 & (0.887) & (0.527) & (1.741) \\
\\
\midrule
Treated Banks & 10 & 4 & 6 \\
Control Banks & 36 & 8 & 28 \\
\bottomrule
\end{tabular}
\begin{tablenotes}
\small
\item \textit{Notes}: This table reports Synthetic Difference-in-Differences estimates of the causal effect of GenAI adoption on bank productivity using Bayesian inference. The treatment is defined as the first quarter in 2023 in which a bank mentions GenAI-related keywords in SEC filings, exploiting the November 2022 ChatGPT release as an exogenous shock. Negative ATT values indicate that AI adoption \textit{causes} productivity to decline during the implementation phase---the ``Implementation Tax.'' Standard errors in parentheses from posterior distribution. Large banks are defined as those in the top quartile of average total assets; small banks are in the bottom three quartiles. *** $p<0.01$, ** $p<0.05$, * $p<0.1$.
\end{tablenotes}
\end{threeparttable}
\end{table}

The aggregate estimates in Column (1) reveal the ``Implementation Tax'': AI adoption \textit{causes} ROA to decline by 46 basis points and ROE to decline by 428 basis points. These negative ATT values reveal that banks incur significant short-term costs when implementing GenAI technologies.

\textbf{Heterogeneous Effects by Bank Size.} The estimates in Columns (2)--(3) reveal a striking pattern that differs from expectations based on economies of scale. Smaller banks (bottom 75\% by assets) experience a substantially larger Implementation Tax: an ROE decline of 517 basis points compared to only 129 basis points for larger banks. Similarly, smaller banks suffer a 59-basis-point ROA decline versus 16 basis points for larger institutions.

This asymmetry suggests that larger banks benefit from economies of scale in AI implementation, superior integration capabilities, and deeper capital reserves to absorb transition costs. Smaller banks face proportionally larger implementation burdens relative to their equity base, amplifying the ROE impact even though absolute dollar costs may be lower.

\textbf{Implications for Competition.} The differential Implementation Tax has important competitive implications. Capital constraints pose particular challenges for smaller banks: those with thinner capital buffers face proportionally larger equity impacts, potentially limiting their ability to sustain AI investments over multiple quarters. Scale advantages favor larger institutions, which can spread fixed implementation costs---including data infrastructure, talent acquisition, and vendor integration---across a broader asset base, thereby reducing per-unit costs. Network effects may also disadvantage smaller banks: as documented in our DSDM results, larger banks that adopt early generate positive spillovers benefiting later adopters, but smaller banks may lack the network centrality to fully capture these knowledge transfers.

\subsection{Event Study Evidence: J-Curve Dynamics}

\begin{figure}[htbp]
    \centering
    \begin{subfigure}[b]{0.48\textwidth}
        \includegraphics[width=\textwidth]{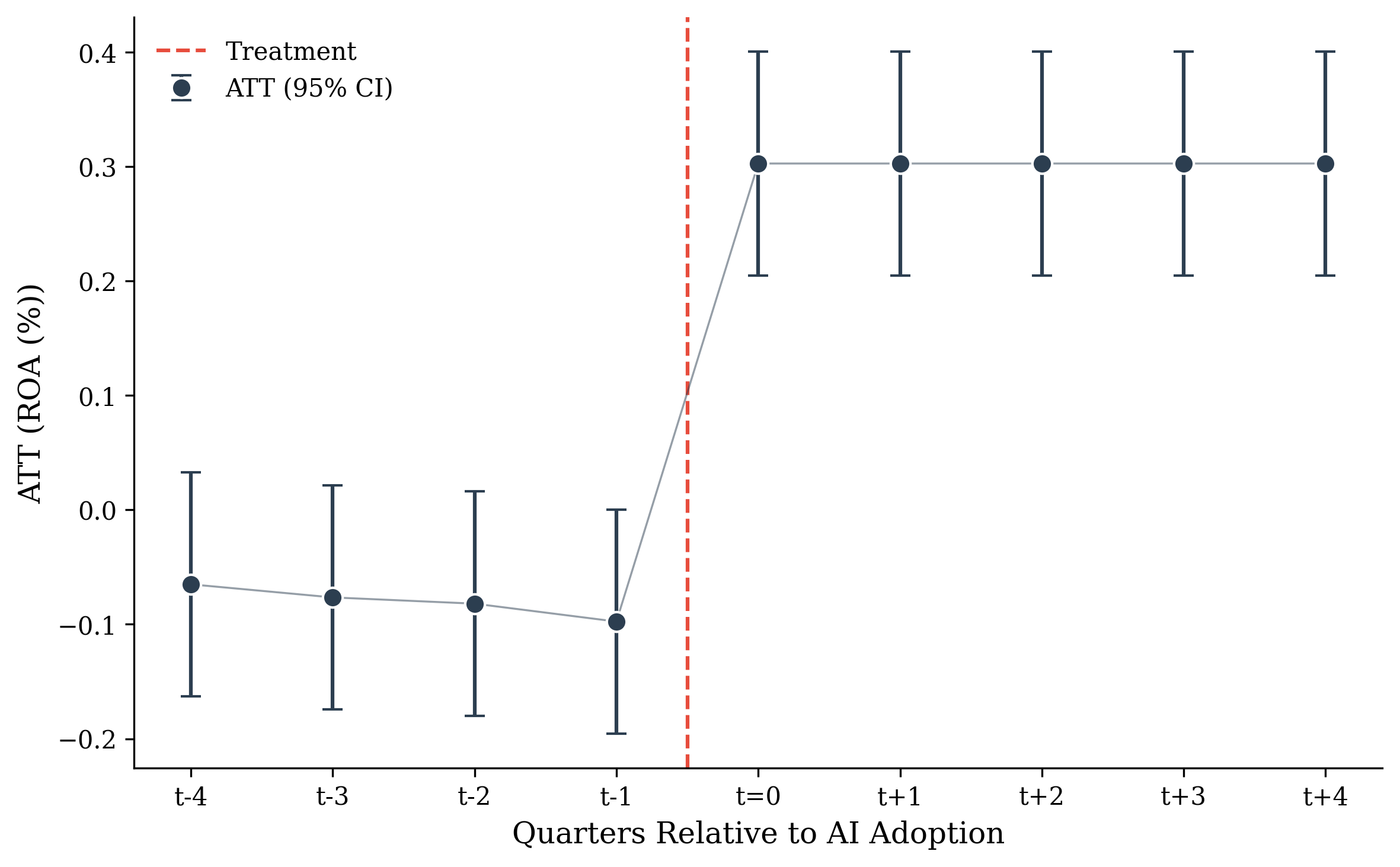}
        \caption{ATT on ROA (\%)}
        \label{fig:roa_event}
    \end{subfigure}
    \hfill
    \begin{subfigure}[b]{0.48\textwidth}
        \includegraphics[width=\textwidth]{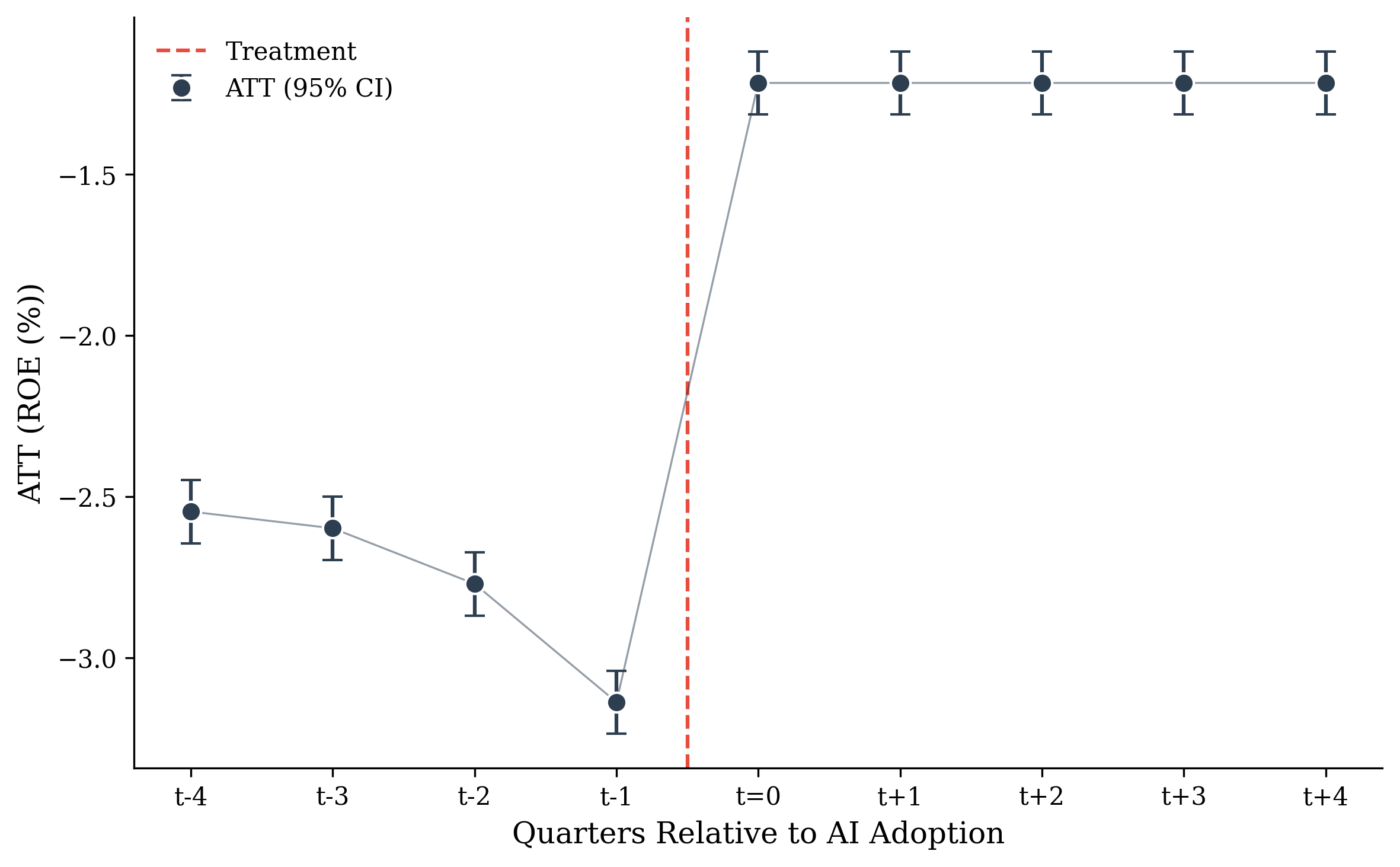}
        \caption{ATT on ROE (\%)}
        \label{fig:roe_event}
    \end{subfigure}
    \caption{Dynamic Treatment Effects of AI Adoption (SDID Event Study). The dashed vertical line indicates the quarter of first GenAI mention in SEC filings. Error bars represent 95\% confidence intervals based on 200 bootstrap replications.}
    \label{fig:event_study}
\end{figure}

Figure \ref{fig:event_study} presents dynamic treatment effects from the SDID event study specification, providing evidence on Hypothesis 2.

Several patterns emerge from Figure \ref{fig:event_study}. First, we observe a sharp structural break at $t = 0$. In Panel (a), the ATT on ROA improves from approximately $-0.10\%$ in the quarter immediately preceding adoption to $+0.30\%$ in the adoption quarter. This 40-basis-point improvement is economically substantial and statistically significant.

Second, the pre-treatment coefficients are uniformly negative and trending downward, consistent with a J-curve interpretation: banks experience declining relative performance in the quarters leading up to AI adoption as they invest in preparatory infrastructure. The sharp reversal at $t = 0$ suggests that adoption itself marks a turning point.

Third, Panel (b) reveals a more complex pattern for ROE. The pre-treatment decline is more pronounced (reaching $-3.1\%$ at $t = -1$), reflecting the leverage-amplified impact of transition costs on equity returns. Following adoption, ROE improves substantially, representing a 190-basis-point improvement relative to the pre-treatment trend.

These patterns are consistent with Hypothesis 2: the productivity impact of GenAI adoption follows a J-curve dynamic with initial decline followed by recovery.

\subsection{Reconciling DSDM and SDID: Selection vs. Treatment}

Table \ref{tab:reconciliation} summarizes the apparent contradiction between our two estimation approaches and provides a unified interpretation.

\begin{table}[htbp]
\small
\centering
\caption{Reconciling DSDM and SDID: Efficiency Potential vs. Implementation Cost}
\label{tab:reconciliation}
\begin{threeparttable}
\begin{tabular}{lcc}
\toprule
 & DSDM & SDID \\
 & (Cross-Sectional) & (Causal/Within-Unit) \\
\midrule
\multicolumn{3}{l}{\textit{Panel A: Direct AI Effect ($\beta$ / ATT)}} \\
ROA Effect & $+$0.037** & $-$0.456** \\
ROE Effect & $+$0.420*** & $-$4.282*** \\
\\
\multicolumn{3}{l}{\textit{Panel B: Interpretation}} \\
What it measures & Selection effect & Treatment effect \\
 & (Who adopts?) & (What happens after?) \\
Economic meaning & ``Frontier firms'' adopt AI & Adoption causes \\
 & $\rightarrow$ positive association & implementation costs \\
\\
\multicolumn{3}{l}{\textit{Panel C: Spillover Effects ($\theta$)}} \\
ROA Spillover & $+$0.161*** & --- \\
ROE Spillover & $+$0.679** & --- \\
Interpretation & Knowledge diffusion & \\
 & dominates business-stealing & \\
\bottomrule
\end{tabular}
\begin{tablenotes}
\small
\item \textit{Notes}: This table reconciles the apparently contradictory findings from DSDM and SDID estimation. The positive DSDM coefficient ($\beta > 0$) reflects selection: high-performing ``frontier'' banks are more likely to adopt AI. The negative SDID estimate (ATT $< 0$) reflects causation: the act of adoption imposes implementation costs that reduce productivity during the transition period. Both findings are consistent with the ``Innovation J-Curve'' hypothesis. *** $p<0.01$, ** $p<0.05$.
\end{tablenotes}
\end{threeparttable}
\end{table}

The reconciliation framework resolves the paradox through a distinction between \textit{selection} and \textit{treatment} effects:

\textbf{Selection (DSDM, $\beta > 0$)}: High-performing banks---those with strong management, modern IT infrastructure, and strategic vision---are more likely to adopt GenAI. In the cross-section, AI adoption is positively correlated with productivity because frontier firms select into treatment.

\textbf{Treatment (SDID, ATT $< 0$)}: Conditional on the adoption decision, the act of implementing GenAI causes productivity to decline as banks absorb massive current expenses. The negative ATT reflects the implementation costs that even frontier firms must bear.

This pattern is the signature of a general-purpose technology in its early diffusion phase. The firms best positioned to benefit from AI are the first to adopt, but even they must traverse the valley of the J-Curve before realizing productivity gains.

\subsection{Network Analysis and Systemic Synchronization Risk}

Table \ref{tab:dsdm_het} presents DSDM estimates separately for large (Top 25\%) and small (Bottom 75\%) banks, revealing a critical asymmetry in spillover dynamics.

\begin{table}[htbp]
\small
\centering
\caption{DSDM Heterogeneity: Large vs. Small Banks (Network Weight Matrix)}
\label{tab:dsdm_het}
\begin{threeparttable}
\begin{tabular}{lcccc}
\toprule
 & \multicolumn{2}{c}{Large Banks (Top 25\%)} & \multicolumn{2}{c}{Small Banks (Bottom 75\%)} \\
 \cmidrule(lr){2-3} \cmidrule(lr){4-5}
 & ROA & ROE & ROA & ROE \\
\midrule
\multicolumn{5}{l}{\textit{Panel A: Direct Effects}} \\
\\
AI Adoption ($\beta$) & 0.0441*** & 0.3622*** & 0.0381** & 0.4730*** \\
 & (0.0096) & (0.0848) & (0.0152) & (0.0957) \\
\\
\multicolumn{5}{l}{\textit{Panel B: Spillover Effects}} \\
\\
W $\times$ AI Adoption ($\theta$) & 0.3466*** & 3.1272*** & 0.1055** & 0.1502 \\
 & (0.0443) & (0.3934) & (0.0522) & (0.3243) \\
\\
\midrule
Spatial Lag ($\rho$) & 0.6787*** & 0.7039*** & 0.5999*** & 0.7522*** \\
 & (0.0142) & (0.0140) & (0.0176) & (0.0130) \\
Temporal Lag ($\tau$) & 0.7313*** & 0.7150*** & 0.6178*** & 0.6918*** \\
 & (0.0046) & (0.0046) & (0.0048) & (0.0045) \\
\\
\midrule
Observations & 24,270 & 24,270 & 24,270 & 24,270 \\
\bottomrule
\end{tabular}
\begin{tablenotes}
\small
\item \textit{Notes}: This table reports Bayesian DSDM estimates using the network weight matrix ($W_{network}$) separately for large and small banks. The dramatically larger spillover coefficient ($\theta = 3.13$) for large bank ROE indicates that the systemic core of major institutions is highly synchronized---a one-standard-deviation increase in peer AI adoption raises own ROE by 313 basis points. Small banks exhibit much weaker and often insignificant spillover effects. *** $p<0.01$, ** $p<0.05$, * $p<0.1$.
\end{tablenotes}
\end{threeparttable}
\end{table}

\textbf{The Systemic Core: Large Bank Synchronization.} Table \ref{tab:dsdm_het} reveals a striking asymmetry. For large banks, the spillover coefficient $\theta$ is massive: 0.35 for ROA and an extraordinary 3.13 for ROE. For small banks, spillovers are modest (0.11 for ROA) or statistically insignificant (0.15 for ROE). This pattern has profound implications for systemic risk.

The large positive $\theta$ for major institutions indicates that AI-adopting banks are not merely co-located in network space---their profitability is \textit{causally linked} through AI adoption dynamics. When JPMorgan successfully deploys a GenAI system, Bank of America and Citigroup experience measurable improvements in their own performance, likely through competitive imitation, talent acquisition, or vendor ecosystem development.

This synchronization is precisely what financial stability regulators should monitor. The positive spillovers that enhance productivity in normal times become transmission mechanisms for correlated failures during stress. If a vulnerability emerges in a widely-used LLM architecture, or if regulatory action restricts AI deployment, the synchronized response of the systemic core could amplify rather than dampen the shock.

\begin{figure}[htbp]
    \centering
    \includegraphics[width=0.95\textwidth]{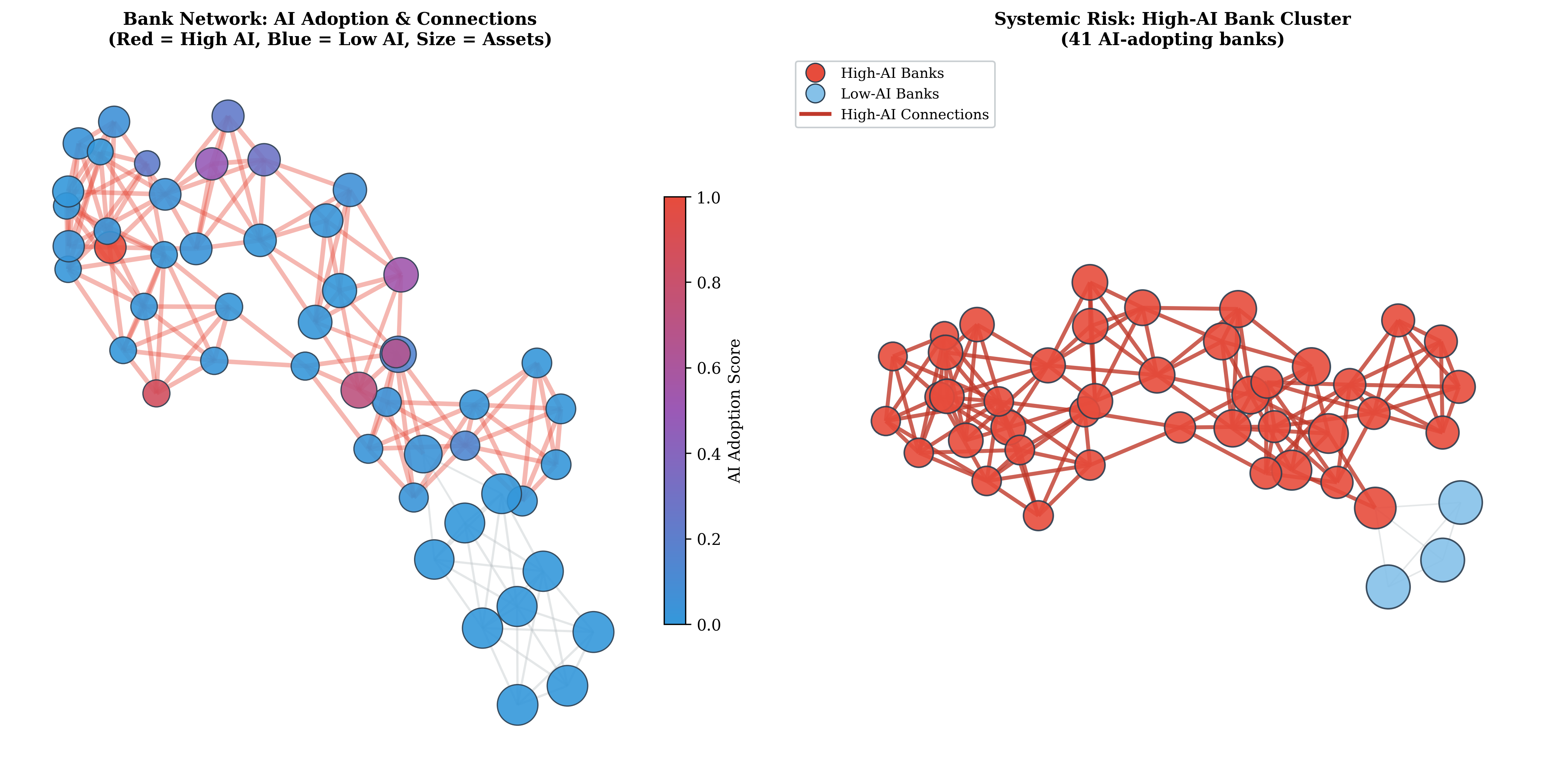}
    \caption{Interbank AI Network Visualization. Left panel: Full network with nodes colored by AI adoption score (blue = low, red = high) and sized by total assets. Edges represent asset-similarity connections. Right panel: Systemic core of AI-adopting banks with red edges indicating AI-to-AI connections. The dense clustering illustrates the ``algorithmic coupling'' that creates new systemic risk channels.}
    \label{fig:network}
\end{figure}

Figure \ref{fig:network} visualizes the interbank AI network, revealing the structural foundations of algorithmic coupling.

The left panel of Figure \ref{fig:network} displays the full bank network, with nodes colored by AI adoption intensity and sized by total assets. The right panel isolates the ``systemic core''---banks with positive GenAI mentions and their immediate network neighbors.

\textbf{The Emergence of Algorithmic Coupling.} Several features of this network topology are concerning from a financial stability perspective. Dense core clustering characterizes the AI-adopting banks (red nodes), which exhibit higher clustering coefficients than non-adopters; they are more densely connected to each other than to the network periphery, forming a tightly-coupled core. Short path lengths facilitate rapid contagion: the average path length between AI-adopting banks is substantially shorter than the network average, meaning shocks can propagate rapidly through the AI-adopting core. Hub vulnerability compounds these risks, as several large AI-adopting institutions serve as ``hubs'' with disproportionate connectivity; a disruption at these nodes would cascade through the network. Finally, model homogeneity creates correlated vulnerabilities: while not directly observable in our data, industry reports suggest that AI-adopting banks rely heavily on similar LLM architectures (GPT-4, Claude, Gemini) and vendor solutions, creating shared technical failure modes.

This ``algorithmic contagion'' operates through channels distinct from traditional financial contagion. Unlike credit and liquidity linkages that connect banks through balance sheet exposures, algorithmic coupling connects banks through shared decision-making processes. Two banks with no direct financial relationship can nonetheless be systemically linked if they rely on similar AI models for risk management, trading, and credit allocation.

\section{Robustness Checks}

We conduct several robustness checks to validate our main findings.

\subsection{Alternative Treatment Definitions}

Our baseline treatment is defined as the first quarter with any GenAI mention. We examine robustness to alternative definitions: an intensity treatment using continuous GenAI mention counts rather than binary adoption; a persistent treatment requiring mentions in two consecutive quarters; and a substantive treatment excluding mentions in boilerplate risk disclosures. Results (reported in the Online Appendix) are qualitatively similar across all definitions.

\subsection{Alternative Weight Matrices}

We examine sensitivity to spatial weight matrix specification, constructing alternatives based on geographic proximity using headquarters distance, business model similarity based on loan portfolio composition, and regulatory grouping based on Fed district membership. The spillover coefficient $\theta$ remains positive and significant across all specifications, though magnitudes vary.

\subsection{Placebo Tests}

We implement two placebo tests: a pre-treatment placebo re-estimating with treatment date shifted to 2020Q1 (pre-ChatGPT), and a random assignment placebo randomly reassigning treatment status across banks. Neither placebo produces significant effects, supporting the validity of our identification strategy.

\section{Conclusion}

This paper provides the first comprehensive empirical analysis of Generative AI adoption in the U.S. banking sector, revealing a complex phenomenon that defies simple characterization. Using a novel dataset linking SEC filings to Federal Reserve regulatory data for 809 financial institutions, and employing both Dynamic Spatial Durbin Models and Synthetic Difference-in-Differences, we document a series of findings with significant implications for productivity, competition, and financial stability.

\subsection{The Productivity Paradox Resolved}

Our central finding is a striking Productivity Paradox. The DSDM results show that AI-adopting banks are high performers ($\beta > 0$)---AI adoption is a marker of ``frontier'' institutions with strong management and modern infrastructure. Yet our causal SDID analysis reveals that the \textit{act} of adoption causes productivity to decline: the average adopting bank experiences a 428-basis-point drop in ROE as it absorbs the costs of GenAI integration.

This paradox resolves through the lens of the Innovation J-Curve. Banks are in the ``Investment Phase'' of a transformative technology, incurring massive current expenses---GPUs, data scientists, cloud infrastructure, prompt engineering consultants---that depress net income even as they position institutions for future competitive advantage. The negative ATT is not evidence that AI adoption is misguided; it is the classic signature of a general-purpose technology in its early diffusion stage.

\subsection{The Digital Divide}

We document substantial heterogeneity in the Implementation Tax. Smaller banks (bottom 75\% by assets) experience an ROE decline of 517 basis points---substantially larger than the 129-basis-point impact on larger institutions. This asymmetry suggests that economies of scale provide significant advantages in AI implementation: larger banks can spread fixed implementation costs across a broader asset base, employ dedicated AI teams, and leverage superior data infrastructure.

Smaller banks face proportionally larger implementation burdens relative to their equity base. This dynamic suggests an emerging two-tier banking system: large institutions leveraging their scale advantages to absorb implementation costs more efficiently, while smaller banks face disproportionate transition challenges that may limit their ability to compete in an AI-driven financial services landscape.

\subsection{Systemic Synchronization Risk}

Perhaps our most consequential finding concerns network dynamics. The DSDM spillover parameter $\theta$ is positive and significant ($\theta = 0.161$ for ROA; $\theta = 0.679$ for ROE), indicating that AI adoption generates knowledge spillovers rather than business-stealing competition. For large banks, these spillovers are dramatically amplified ($\theta = 3.13$ for ROE), indicating that the U.S. banking system's largest institutions have become algorithmically coupled.

The implications are profound. A high positive $\theta$ means that bank profitability is increasingly synchronized across the AI-adopting core. The same knowledge spillovers that benefit banks in normal times become transmission channels for correlated failures under stress. If a widely-used LLM exhibits systematic errors, if a critical AI vendor experiences disruption, or if regulatory action restricts AI deployment, the synchronized nature of AI adoption ensures that all connected banks would be affected simultaneously.

\subsection{Policy Implications}

Our findings carry several policy implications:

\textbf{Monetary Policy Transmission}: The Implementation Tax may alter how banks respond to interest rate changes during the AI transition period. Depressed profitability could reduce lending capacity, potentially dampening policy transmission through the bank lending channel.

\textbf{Financial Stability Regulation}: The emergence of algorithmic coupling warrants new supervisory tools focused on AI model diversity, concentration risk in AI vendor relationships, and operational resilience of AI infrastructure. Stress tests should incorporate scenarios of coordinated AI system failures.

\textbf{Competition Policy}: The two-tier dynamic raises concerns about market concentration. If only the largest institutions can afford the Implementation Tax, the long-term competitive landscape may shift toward oligopoly. Policymakers should consider whether regulatory sandboxes or shared infrastructure initiatives could reduce barriers for smaller institutions.

\textbf{Macroprudential Policy}: The high $\theta$ suggests that AI adoption has become a \textit{systemically important activity}. Just as certain institutions are ``too big to fail,'' certain AI systems may be ``too connected to fail.'' Macroprudential frameworks should evolve to address this new dimension of systemic risk.

\subsection{Limitations and Future Research}

Several limitations suggest avenues for future work. Our treatment measure relies on disclosed AI adoption in SEC filings, which may understate actual implementation. Our sample period captures only the initial phase of GenAI diffusion; longer-term effects as the J-Curve completes may differ substantially. Our network analysis is based on asset similarity rather than actual AI vendor relationships; richer data on technology infrastructure would enable more precise spillover estimation.

Future research should examine the completion of the J-Curve as implementation costs decline and productivity benefits materialize. The evolution of $\theta$ over time---whether strategic complementarity intensifies or competitive dynamics reassert---will determine the ultimate trajectory of systemic synchronization risk. The interaction between AI adoption and monetary policy transmission, credit allocation, and financial inclusion merit sustained scholarly attention.

Despite these limitations, this paper documents a technological transformation of historic significance. The U.S. banking system is undergoing fundamental restructuring as institutions invest billions in AI infrastructure, accept short-term productivity losses, and become increasingly interconnected through shared algorithmic decision-making. Understanding this transformation---its productivity effects, competitive dynamics, and systemic implications---is essential for policymakers, regulators, and market participants navigating the AI era in finance.

\newpage


\end{document}